\documentclass[prl,twocolumn]{revtex4}
\usepackage{amssymb,amsmath,amsthm,graphicx,psfrag}

\pdfoutput=1

\begin{document}


\def\be{\begin{equation}}
\def\ee{\end{equation}}
\def\bea{\begin{eqnarray}}
\def\eea{\end{eqnarray}}
\newcommand{\der}[2]{\frac{\partial{#1}}{\partial{#2}}}
\newcommand{\dder}[2]{\partial{}^2 #1 \over {\partial{#2}}^2}
\newcommand{\dderf}[3]{\partial{}^2 #1 \over {\partial{#2} \partial{#3}}}
\newcommand{\eq}[1]{Eq.~(\ref{eq:#1})}
\newcommand{\dd}{\mathrm{d}}

\title{Semiclassical instabilities of Kerr-AdS black holes}

\vskip 4cm

\author{R. Monteiro}
\email{R.J.F.Monteiro@damtp.cam.ac.uk}
\author{M. J. Perry}
\email{M.J.Perry@damtp.cam.ac.uk}
\author{J. E. Santos}
\email{J.E.Santos@damtp.cam.ac.uk}

\vskip 0.2cm

\affiliation{DAMTP, Centre for Mathematical Sciences, University of Cambridge, Wilberforce Road, Cambridge CB3 0WA, UK}

\vskip 0.5cm

\date{\today}

\vskip 4cm

\begin{abstract}
We study the thermodynamic stability of the Kerr-AdS black hole from the perturbative corrections to the gravitational partition function. The line of critical stability is identified by the appearance of a negative mode of the Euclidean action that renders the partition function ill-defined. The eigenvalue problem, consisting of a system of three coupled partial differential equations for the metric perturbations, is solved numerically. The agreement with the standard condition of thermodynamic stability in the grand canonical ensemble is remarkable. The results illustrate the physical significance of gravitational partition functions for rotating spacetimes beyond the instanton approximation.
At a classical level, the results also imply that the Gregory-Laflamme instability of the Kerr string persists up to extremality, the range of unstable modes increasing with the angular momentum.

\vskip 0.5cm
 
\end{abstract}

\maketitle


\section{Introduction}

The thermodynamics of black holes has been a cornerstone in the search for a quantum theory of gravity. Its semiclassical results, such as the Hawking temperature and the Bekenstein-Hawking entropy, are expected to be reproduced by any serious candidate theory. Indeed their reproduction has been hailed as a major success of string theory, eg. \cite{Strominger:1996sh}. Another influential result of the imaginary time methods was the construction of gravitational partition functions by Gibbons and Hawking \cite{Gibbons:1976ue}. These are defined as path integrals in the same spirit as the partition functions of flat space thermal field theory, but treating the metric as a quantum field, that is
\be
\label{pathintegral}
Z = \int \mathrm{D}[g] e^{-I[g]}.
\ee
It should be emphasised that the gravitational path integral is known to be non-renormalisable and is considered here as a low energy effective theory that may provide clues to a proper theory of quantum gravity.

The boundary conditions of the path integral, and the corresponding boundary terms in the gravitational action $I[g]$, specify the thermodynamic ensemble described by the partition function. The path integral is approximated by an instanton saddle-point $Z \approx e^{-I[\hat{g}]}$, which gives the leading contribution to the free energy of the system. However, the quantum corrections play an important role too - they identify the thermodynamic instabilities of the ensemble by causing a divergence in the path integral. This type of pathology was identified by Gross, Perry and Yaffe \cite{Gross:1982cv} for the Schwarzschild black hole, which has a negative specific heat. Furthermore, Prestidge \cite{Prestidge:1999uq} showed that the partition function of the Schwarzschild-AdS black hole becomes well-defined for large black holes exactly when the specific heat becomes positive. The inclusion of matter presents technical difficulties but the same rule holds for Reissner-Nordstr\"om black holes \cite{Monteiro:2008wr}. An argument in support of the rule was given by Reall \cite{Reall:2001ag}.

In this paper, we deal with the one-loop quantum corrections to the partition functions of Kerr-AdS black holes. In particular, we look for negative modes of the second order gravitational action. This is a long-standing problem, even in the asymptotically flat case, for two reasons. First, the imaginary time methods seem subtler since the instanton metrics for rotating black holes are complex, or quasi-Euclidean, rather than real with Euclidean signature. The instanton is just the leading order contribution to the path integral and it has a real action \cite{Brown:1990di,Brown:1990fk}. The second difficulty is that the lack of symmetry of the problem makes it much harder to solve than the static cases. We address this with an improved numerical method to solve coupled partial differential equations, and by concentrating on the space of perturbative corrections likely to identify an instability. The technique in this work differs from the simple but incomplete first approach to the problem in \cite{Monteiro:2009tc}, which could only account for the effect of a single direction in the perturbation space.

The first law of thermodynamics for a vacuum black hole with mass $M$ and angular momentum $J$ is
\be
\dd M= T \dd S + \Omega \dd J.
\ee
In the grand canonical ensemble, the temperature $T$ and the angular velocity $\Omega$ are fixed, and the important thermodynamic potential is the Gibbs free energy $G=M-T S-\Omega J$. Standard arguments \cite{Landau} lead to the stability condition for equilibrium: the Hessian matrix
\be
{\dderf{(-G)}{y_\mu}{y_\nu}}, \qquad y_\mu=(T,\Omega),
\ee
must be positive definite. The two eigenvalues of this matrix are $\beta C_J = (\partial{S} / \partial{T})_J$, where $\beta$ is the inverse temperature and $C_J$ is the specific heat at constant angular momentum; and the isothermal moment of inertia $\epsilon_T = (\partial{J} / \partial{\Omega})_T$. In the canonical ensemble, the positivity of $C_J$ is sufficient to ensure perturbative stability. In the grand-canonical ensemble however, the moment of inertia $\epsilon_T$ is also required to be positive, and together with $C_J$ renders asymptotically flat black holes thermodynamically unstable \cite{Monteiro:2009tc}. We analyse here the Kerr-AdS black hole not only because of the usual theoretical motivations (eg. the AdS/CFT correspondence \cite{Maldacena:1997re}) but also because there is a phase transition to stability for large enough black holes in the grand-canonical ensemble. Our purpose is to compare this criterion for stability with the one given by the quantum corrections to the gravitational partition function.

Let us describe now the path integral and the quantum corrections. The Euclidean gravitational action in the path integral (\ref{pathintegral}) is given by
\be
\label{action}
I[g] = -{1 \over 16 \pi} \int_{\mathcal M} \dd V\, (R-2 \Lambda) -{1 \over 8 \pi} \int_{\partial {\mathcal M}} \dd \Sigma\, K - I_0,
\ee
where the last term is a background subtraction that normalises the action of pure AdS space to zero (see also its holographic interpretation \cite{Balasubramanian:1999re,Skenderis:2000in,Olea:2005gb,Olea:2006vd}). The second term is the York-Gibbons-Hawking boundary term \cite{York:1972sj,Gibbons:1976ue}, which is appropriate for boundary conditions of fixed induced metric on ${\partial {\mathcal M}}$. This corresponds to the grand-canonical ensemble, since fixing instead the charges $M$ or $J$ would require fixing derivatives of the metric normal to the boundary. The instanton $\hat{g}$ is a non-singular solution to the equations of motion derived from this action, and the quantum corrections are given by off-shell perturbations about the instanton solution,
\be
g_{ab} = \hat{g}_{ab} + h_{ab}.
\ee
Pure trace $h_{ab} = \hat{g}_{ab}\, h^c_{\phantom{c}c} /4$ (conformal) metric perturbations render the action unbounded below but can be shown to decouple and give no contribution to the partition function, at least at the one-loop perturbative level \cite{Gibbons:1978ac,Gibbons:1978ji}. This is the timelike direction in the Wheeler-DeWitt metric. The gauge-invariant perturbations are the traceless-transverse (TT) modes, such that ${h^{TT}}^a_{\phantom{a}a}=0$ and $\nabla^a h_{ab}^{TT}=0$. Their contribution to the one-loop path integral is given by
\be
\label{piTT}
\int \mathrm{D}[h^{TT}] \; \exp{\left( - \frac{1}{64 \pi} \int \dd V\, h^{TT} \cdot G h^{TT}\right)},
\ee 
where the operator $G$ is defined as
\be
(G h)_{ab} = - \nabla^c \nabla_c h_{ab} -2 R_{a\phantom{c}b}^{\phantom{a}c\phantom{b}d} h_{cd}.
\ee 
All metric operations hereafter are made with respect to the instanton background metric $\hat{g}_{ab}$. It is clear from the expression (\ref{piTT}) that if the spectrum of the operator $G$ on TT modes possesses a negative eigenvalue, the perturbative corrections are ill-defined (zero modes can be dealt with by the standard collective coordinates method \cite{Jevicki:1976kd}).

\section{The eigenvalue problem}

We will look at eigenvalues of the operator $G$ in search for negative modes, $G h^{TT} =  \lambda \, h^{TT}$, with $\lambda <0$. In the static spherically symmetric examples, the unique negative eigenmode is radial as if it changed the `mass'. For the rotating black holes, we will consider only stationary axisymmetric perturbations, so that the isometries of the background are preserved.

Our ansatz for the perturbed metric is given by
\bea
\dd s^2 = \frac{\Delta}{\Sigma^2} e^{\alpha}  \left( \dd\tau -i \frac{a}{\Xi} e^{\omega} \sin^2\theta \dd\phi \right)^2 + \nonumber \\
+ \frac{\Delta_\theta \sin^2\theta}{\Sigma^2} e^{\sigma}  \left( \frac{r^2+a^2}{\Xi} \dd\phi + i a e^{-\omega} \dd\tau \right)^2+ \nonumber \\
+ \frac{\Sigma^2}{\Delta} e^{\gamma} (\dd r + r_0 \chi \sin \theta \dd \theta )^2 +  \nonumber\\
+ \frac{\Sigma^2}{\Delta_\theta} e^{\eta} \dd\theta^2,
\label{eq:kerrAdS1}
\eea
where $\ell$ is the curvature radius of AdS and is related to the cosmological constant as $\ell^2 = - 3/\Lambda$, and
\begin{subequations}
\begin{eqnarray}
\Delta (r) &=& (r^2+a^2)(1+r^2\ell^{-2}) -r_0 r, \\
\Sigma (r,\theta)^2 &=& r^2 +a^2 \cos^2\theta, \\
\Delta_\theta (\theta) &=& 1-a^2\ell^{-2} \cos^2{\theta}, \\
\Xi &=& 1-a^2\ell^{-2}.
\end{eqnarray}
\end{subequations}
The functions $\alpha$, $\omega$, $\sigma$, $\gamma$, $\chi$, $\eta$ are small perturbations and depend on both $r$ and $\theta$. If all vanish, the line element above represents the Kerr-AdS instanton with mass $M=r_0/2$. The bounds on the parameter space of the instanton are given by the extremality condition and by the requirement that $|a|<\ell$, since the limit $|a| \to \ell$ is singular \cite{Hawking:1998kw}. Moreover, the avoidance of a conical singularity at the instanton horizon, located at $r= r_{_+}$, the largest root of $\Delta$, requires the coordinate identification $(\tau,\phi)= (\tau+ \beta, \phi - i \beta (\Omega -a / \ell^{-2}))$. Here,
\be
\beta=\frac{4\pi(r_{_+}^2 +a^2)}{r_{_+}(1+a^2\ell^{-2} + 3 r_{_+}^2 \ell^{-2} - a^2r_{_+}^{-2})}
\ee
is the inverse temperature and
\be
\Omega = \frac{a(1+ r_{_+}^2\ell^{-2})}{r_{_+}^2 + a^2}
\ee
is the angular velocity in a reference frame non-rotating at infinity, the quantity that satisfies the first law of thermodynamics \cite{Gibbons:2004ai}.

The imposition of the TT conditions on the first order perturbation $h_{ab}$ implies that $\alpha$, $\omega$, $\sigma$ are given by $\gamma$, $\chi$, $\eta$ and their first derivatives. Substituting these relations in the metric perturbation we get the general expression for $h^{TT}_{ab}$ respecting the isometries of the instanton.

Now we look at the eigenvalue problem $G h^{TT} =  \lambda \, h^{TT}$. There are six equations to be solved, corresponding to the components $\tau\tau$, $\tau\phi$, $\phi\phi$, $rr$, $r\theta$, $\theta\theta$. However, the components $\tau\tau$, $\tau\phi$ and $\phi\phi$ are differential equations of third order when we substitute the expressions for $\alpha$, $\omega$, $\sigma$ given by the TT conditions, and they are solved automatically by the remaining three equations as expected. We thus have the three coupled partial differential equations $rr$, $r\theta$ and $\theta\theta$ for the three unknown functions $\gamma$, $\chi$ and $\eta$.

Let us first consider the boundary conditions for the perturbations. The region of integration for the differential equations is an infinite strip $r_+ \leq r < \infty$, $0 \leq \theta \leq \pi$. The perturbations must vanish sufficiently fast at infinity, $r \to \infty$, so that the metric perturbation is normalisable $\int dV h^{TT} \cdot h^{TT} < \infty$. The boundary conditions on the horizon are given by expanding the differential equations around $r=r_+$. We find that this implies $\chi (r_+,\theta) = A / \sin^2 \theta$, $A$ being an integration constant. Regularity for $\theta = 0 ,\pi$ thus requires $\chi (r_+,\theta) = 0$. For $\gamma (r_+,\theta)$ and $\chi (r_+,\theta)$, regularity and normalisability near the horizon require simply that they are finite.

One thing we should emphasise is that the subtlety of having a quasi-Euclidean metric, as opposed to Euclidean, has vanished now. The equations obtained are real equations for real perturbation functions. Unfortunately, it will not be possible to present the three equations explicitly here. Even in the asymptotically flat limit $\ell \to \infty$, they are too cumbersome.

It is convenient to make rescalings such that only adimensional quantities are involved in the problem,
\begin{subequations}
\begin{eqnarray}
y = \frac{r}{r_+} -1, &&  y_\ast = \frac{r_+}{r_0}, \\
\ell_\ast = \frac{\ell}{r_0}, &&  \lambda_\ast = \lambda \, r_0^2,
\end{eqnarray}
and notice that
\be
\frac{y_\ast - y_\ast^2 - y_\ast^4 \ell_\ast^{-2}}{1+ y_\ast^2 \ell_\ast^{-2}} = \left( \frac{a}{r_0} \right)^2. 
\ee
\end{subequations}
We will use the coordinate
\be
x=\cos \theta,
\ee
so that the numerical integration will be performed on a rectangle $0 \leq y \leq Y$, $-1 \leq x \leq 1$, where $Y \gg y_\ast$ must be sufficiently large. Let us rescale the perturbation functions too,
\begin{subequations}
\begin{eqnarray}
\gamma(r,\theta) &=& \frac{q_1(y,x)}{y(1-x^2)} , \\
\chi(r,\theta) &=& \frac{q_2(y,x)}{1-x^2} , \\
\eta(r,\theta) &=& \frac{q_3(y,x)}{y(1-x^2)} ,
\end{eqnarray}
\end{subequations}
so that the boundary conditions discussed before are simply $q_i=0$ ($i=1,2,3$) along the edges $x=\pm 1$ and $y=0,Y$. It is convenient to combine our previous equations $rr$, $r\theta$ and $\theta\theta$, now in terms of the redefined quantities, into the form
\be
\partial^2_y q_i (y,x) + \ldots + \lambda_\ast f(y,x;y_\ast,\ell_\ast) q_i (y,x) =0,
\ee
which turns out to be possible. Notice that the last term above gives the only dependence of the equations on $\lambda_\ast$. We are now ready to implement the spectral numerical method \cite{Trefethen}.

\begin{figure}
\centering
\includegraphics[width = 6 cm]{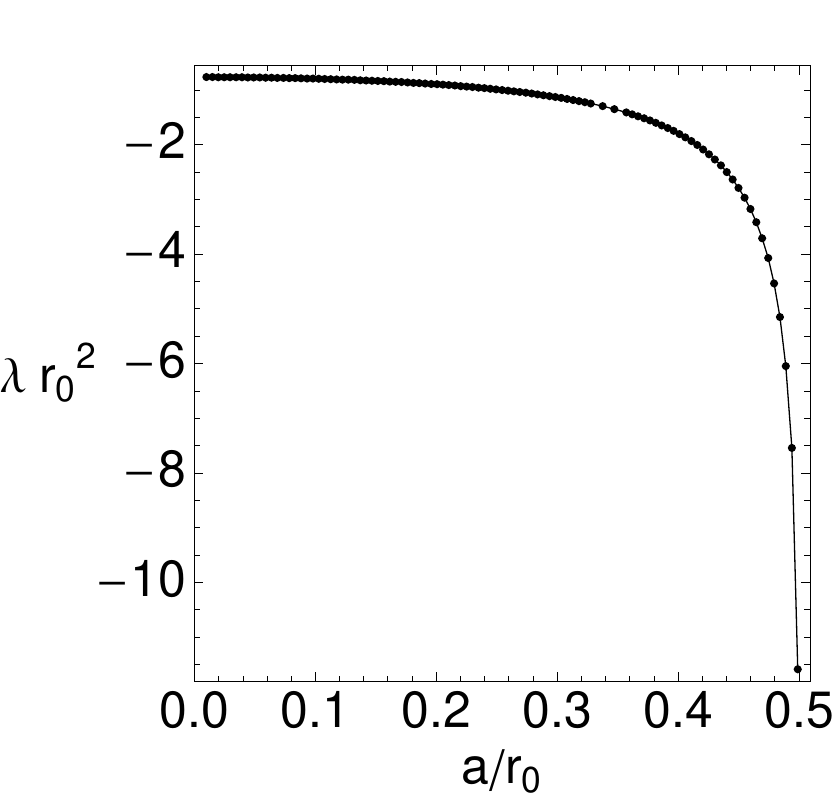}
\caption{\label{fig:kerr}For the Kerr instanton, $\lambda_*$ is negative, decreasing monotonically away from $a=0$ and evaluating to a finite value at extremality $|a| = r_0/2$.}
\end{figure}

Let us first consider the Kerr case, i.e. the limit $\ell \to \infty$, leaving the general case to the next section. The results are represented in Fig.~\ref{fig:kerr}. We find that there is a single negative mode for $|a| \leq r_0/2$, monotonically increasing in magnitude with the angular momentum. Surprisingly, our probe perturbation method \cite{Monteiro:2009tc} approached the value of the negative eigenvalue now found to within 10\%.
One way to understand the increase in magnitude is to recall the connection between the black hole thermodynamic negative mode and the classical Gregory-Laflamme instability of the respective black string/brane \cite{Reall:2001ag}.
The threshold wavenumber $k= (\vec{k} \cdot \vec{k})^{1/2}$ for the Gregory-Laflamme instability corresponds to the four dimensional stationary solution of $G h^{TT} = -k^2 \, h^{TT}$, with the appropriate boundary conditions \cite{Gregory:1993vy}. This is exactly the problem addressed here if we identify $\lambda=-k^2$. The fact that we are dealing with a quasi-Euclidean geometry rather than a Lorentzian geometry is irrelevant since time plays no role in the solutions to the perturbation functions defined in (\ref{eq:kerrAdS1}). The curve in Fig.~\ref{fig:kerr} thus implies that the Gregory-Laflamme instability of the Kerr string persists up to extremality. The larger in magnitude is the negative mode, the smaller is the threshold length scale $k^{-1}$ for the instability. We expect on physical grounds that the centrifugal force caused by the rotation will favour the instability of ripples along the string (Ref.~\cite{Caldarelli:2008mv} presents a fluid dual analogy) thus decreasing their threshold length scale and explaining the stronger negative mode of the black hole. See \cite{Kleihaus:2007dg} for analogous results in higher dimensions with equal angular momenta.

\section{Comparison of stability criteria}

Thermodynamic stability in the grand-canonical ensemble requires that both the specific heat at constant angular momentum $C_J$ and the isothermal moment of inertia $\epsilon_T$ are positive. An interesting feature of the Kerr-AdS black hole, as opposed to the black ring for instance, is that the specific heat at constant angular velocity, $C_\Omega = T (\partial{S} / \partial{T})_\Omega$, is sufficient to describe the stability of the grand-canonical ensemble. It is positive when both $C_J$ and $\epsilon_T$ are positive, and negative when one of them is negative; $C_J$ and $\epsilon_T$ are never simultaneously negative and complement the region of instability. The explicit expression is
\be
\beta C_\Omega= 
-\frac{8 \pi^2  r_+ \left(r_+^2 + a^2\right)}
{\Xi \left(1 + a^2 r_+^{-2} + a^2 \ell^{-2} -3 r_+^2 \ell^{-2} \right)}.
\ee
The vanishing of the denominator in the right-hand-side identifies the line of critical stability for each value of $\ell$. Across this line, $C_\Omega$ diverges and changes sign.

\begin{figure}
\centering
\includegraphics[width = 8 cm]{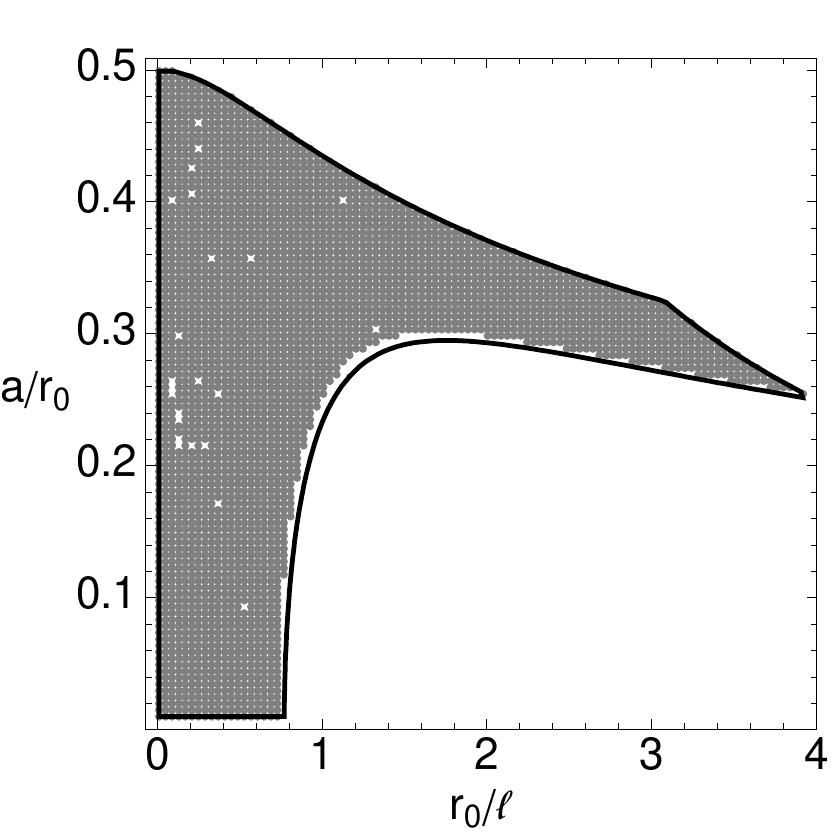}
\caption{\label{fig:kerr_ads}Phase diagram of the Kerr-AdS black hole. The points represent the parameter region with a negative mode and the line represents the change of sign of $C_\Omega$, which is negative in the Kerr limit $\ell \to \infty$.
The diagram is limited above by the extremality bound and by the singular limit $a=l$.}
\end{figure}

If we look at Fig.~\ref{fig:kerr_ads}, the agreement is striking.
There is a single negative mode only when $C_\Omega$ is negative. Unfortunately the numerical method does not allow us to safely zoom in the line of critical stability, so that we cannot make claims about small deviations here. Nevertheless, it is clear that the gravitational partition function reproduces the thermodynamics of the system beyond the instanton approximation, even for this non-static black hole with a quasi-Euclidean instanton.

\section{Acknowledgements}

We are grateful to Gary Gibbons, Stephen Hawking, Gustav Holzegel, Hari Kunduri and Claude Warnick for valuable discussions. RM and JES acknowledge support from the Funda\c c\~ao para a Ci\^encia e Tecnologia (FCT, Portugal) through the grants SFRH/BD/22211/2005 (RM) and SFRH/BD/22058/2005 (JES).

\bibliography{bibliography}

\end{document}